\documentclass[letterpaper]{article} %
\usepackage{aaai2027}  %
\nocopyright  %
\usepackage[hyphens]{url}  %
\usepackage{graphicx} %
\usepackage{natbib}  %
\usepackage{caption} %
\usepackage{amsmath,amssymb,amsfonts}
\usepackage{xcolor}
\usepackage{booktabs}
\usepackage{multirow}
\usepackage{listings}
\usepackage{tikz}
\usepackage{pgfplots}
\pgfplotsset{compat=1.18}
\usepgfplotslibrary{fillbetween}
\usetikzlibrary{arrows.meta, positioning, shapes.geometric, calc, fit, backgrounds}

\usepackage{xspace}
\usepackage[skins]{tcolorbox}
\newtcolorbox{takeaway}{enhanced, size=fbox, left=4mm, right=2mm, boxrule=0pt,
    top=2mm, bottom=2mm, borderline west={2mm}{0pt}{blue!70!black},
    colback=blue!3!white, sharp corners=south}
\newcommand{\sys}{VulAgentRL\xspace}

\usepackage{cleveref}

\begin{document}

\title{Graph Is the Verifier: Agentic Reinforcement Learning for Interprocedural Vulnerability Detection}

\author{
    Yikun Li\textsuperscript{\rm 1},
    Ting Zhang\textsuperscript{\rm 1},
    Jiakun Liu\textsuperscript{\rm 2},
    Jinfeng Jiang\textsuperscript{\rm 1},
    Yieh Yuheng\textsuperscript{\rm 1},
    Yixin Yang\textsuperscript{\rm 1},
    Leow Wen Bin\textsuperscript{\rm 3},
    Yin Yide\textsuperscript{\rm 3},
    Yintong Huo\textsuperscript{\rm 1},
    Eng Lieh Ouh\textsuperscript{\rm 1},
    Lwin Khin Shar\textsuperscript{\rm 1},
    David Lo\textsuperscript{\rm 1}
}
\affiliations{
    \textsuperscript{\rm 1}Singapore Management University, Singapore\\
    \textsuperscript{\rm 2}Harbin Institute of Technology, China\\
    \textsuperscript{\rm 3}GovTech, Singapore\\
    yikunli@smu.edu.sg
}

\maketitle

\begin{abstract}
Real-world vulnerabilities often span multiple functions, yet most learning-based detectors classify each function in isolation: on a sample of real CVEs, we find that 71.7\% of vulnerable functions require evidence from outside the function to be classified correctly. Agentic reinforcement learning (RL) could close this gap by enabling a model to gather that evidence itself, but it lacks a reliable reward, since a reward defined on the final verdict alone can be obtained without performing any investigation. We propose \sys, an agentic RL framework for interprocedural vulnerability detection built on a Code Property Graph (CPG). The CPG serves two roles: at inference time the policy queries it for callers, callees, dataflow, and other queries, and at training time the same graph verifies the evidence the policy cites. Because every CPG node carries a persistent integer identifier, this verification is an exact comparison rather than a textual match, so the reward credits verdicts that are supported by evidence. We further initialize the policy by distilling teacher investigations, and show that this warm start is necessary, since RL cannot acquire tool-use behavior it never samples. Under a repository-level split that prevents leakage, \sys outperforms state-of-the-art baselines, including frontier models, on the strict pair-wise-correct metric while issuing fewer tool calls, and its advantage persists on an out-of-distribution corpus and under class imbalance.
\end{abstract}

\section{Introduction}\label{sec:intro}

\begin{figure}[!t]
\centering
\includegraphics[trim=40 35 40 35, clip, width=\columnwidth]{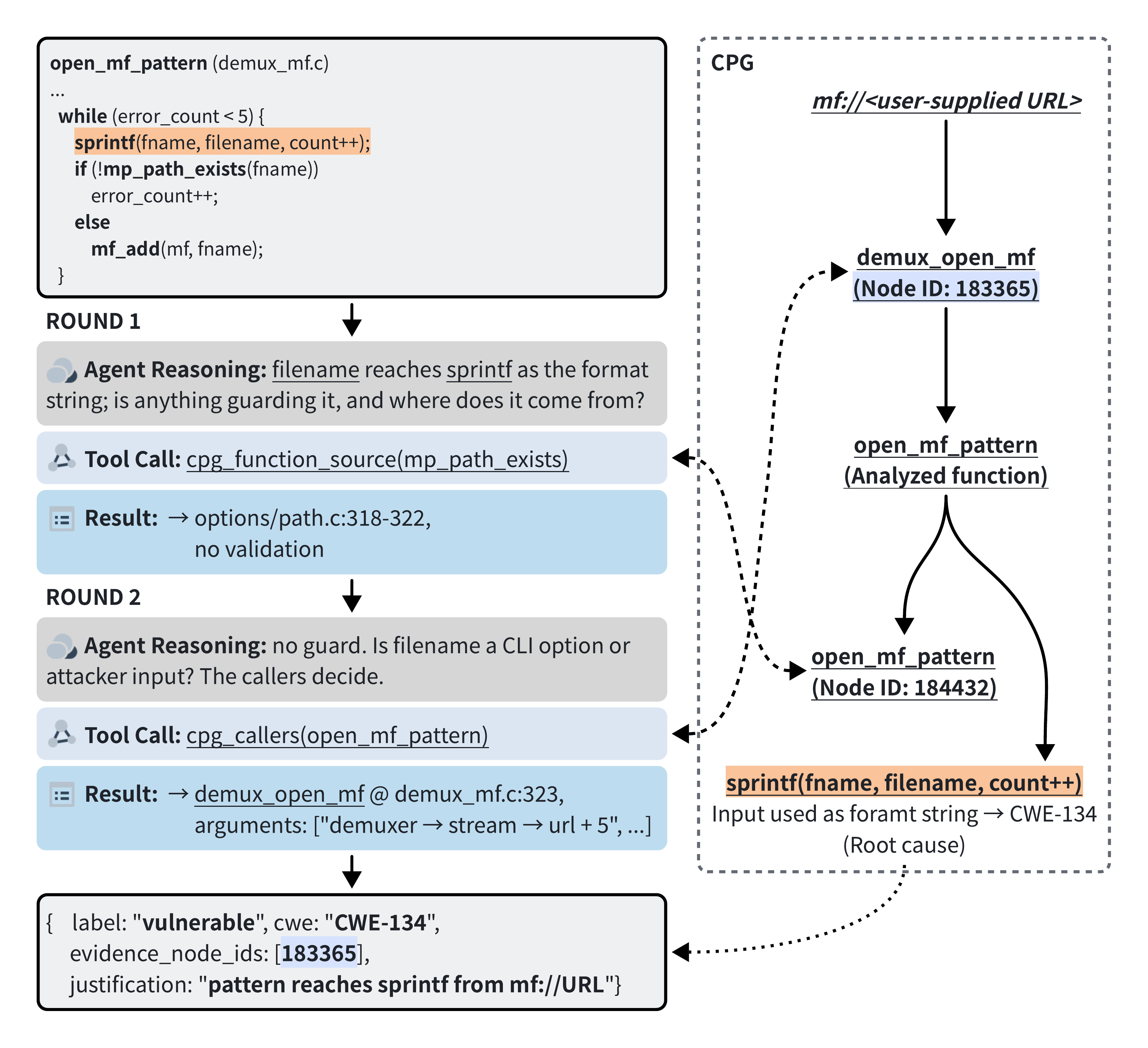}
\caption{\sys investigating a format-string vulnerability in mpv. The agent queries the CPG to rule out a candidate guard, establishes that the format string originates from a user-supplied URL, and cites the graph nodes that justify its verdict.}
\label{fig:teaser}
\vspace{-5mm}
\end{figure}

Software vulnerabilities in widely-deployed code are a persistent source of security risk, and detecting them early, before deployment, is a central goal of secure software engineering. Detection is difficult in practice: real-world vulnerabilities are deeply embedded in large codebases, whether a defect is exploitable typically depends on validation and dataflow conditions scattered across procedures and files, and manual auditing at this scale is expensive and error-prone. Two families of approaches have therefore emerged to automate vulnerability detection. Classical static analyzers~\citep{joern,codeql} encode interprocedural reasoning as hand-crafted dataflow queries: given an appropriate rule, they precisely trace taint from source to sink across functions, but their rule sets are authored per language by experts and scale poorly with the diversity of real-world code. More recently, learning-based detectors~\citep{linevul,vulberta,primevul,titanvul,vultrial} fine-tune or prompt language models to classify a function directly from its source, formulating detection as function-level classification: a model receives a single function and predicts whether it is vulnerable.

Despite this progress, reliable vulnerability detection is impeded by two fundamental challenges: the interprocedural nature of real vulnerabilities and the lack of a verifiable reward for training an investigative agent.

\textbf{First, most existing detectors rely on a single function and cannot reach interprocedural evidence.} The defect and the conditions that determine whether it is exploitable frequently reside in different procedures: on a sample of 138 real vulnerable functions from our corpus, 71.7\% require evidence from at least one caller, callee, or dataflow predecessor outside the function to reach a correct verdict (\Cref{sec:motivation}). A function-level model has no mechanism to fetch such evidence, and inlining the surrounding call graph exceeds practical context windows. Reliable detection therefore requires selective, interprocedural investigation, and a natural remedy is to make the model agentic: equip it with code-navigation tools and train it end-to-end with reinforcement learning (RL)~\citep{grpo}.

\textbf{Second, constructing a reliable reward signal for agentic detection is difficult.} A reward on the final verdict alone is satisfiable without any investigation: policies trained with a verdict-only reward degenerate to evidence-free prediction, exploiting the class prior instead of investigating (\Cref{sec:reward_degeneration}). Execution-free alternatives for supervising the investigation are also inadequate: LLM-as-judge scoring is unreliable in this domain, where judges often fail to distinguish vulnerable from patched functions~\citep{primevul}, and textual matching of cited evidence is brittle under renaming and reformatting. This leaves a critical gap for a reward that verifies what evidence the agent gathered, which in turn requires an evidence representation a reward function can check exactly.

To bridge these gaps, we propose \sys, an agentic RL framework for interprocedural vulnerability detection. The unifying idea is a Code Property Graph (CPG), built once per project with Joern~\citep{joern}, that serves two roles simultaneously: the policy's queryable tool at inference time and the verifier of its evidence at training time.

\noindent 1) \textbf{Agentic CPG querying} (addressing Challenge~1). Rather than classifying the function in isolation, the \sys policy issues typed queries over the CPG and receives structured node sets in return. This enables the model to fetch exactly the interprocedural context it needs (callers, callees, dataflow paths, and other queries) on demand, without inlining the entire call graph into the prompt.

\noindent 2) \textbf{A CPG-as-verifier composite reward} (addressing Challenge~2). \sys is trained in two stages: supervised fine-tuning (SFT) on hint-guided, teacher-distilled multi-turn trajectories, required because GRPO cannot bootstrap tool use from a cold start (\Cref{sec:results}), then Group Relative Policy Optimization (GRPO)~\citep{grpo} under a composite reward over verdict correctness, evidence-node grounding, and CWE classification. Because every CPG node carries a persistent integer ID, the model's cited evidence node IDs are addresses into the same graph the teacher cited from, so the reward verifies evidence by a pure set comparison rather than brittle textual matching. This precise signal enables the reward to distinguish substantive investigation from guessing. \Cref{fig:pipeline} overviews the pipeline.

We evaluate \sys on PrimeVul~\citep{primevul} under a strict repository-level split, adopting the pair-wise-correct rate (P-C) as the primary metric. \sys attains a P-C of 0.378 and an accuracy of 0.633, surpassing the best frontier baseline (Claude Opus 4.7, P-C 0.273) as well as LLM-agent baselines. In particular, it exceeds the state-of-the-art agentic detector JitVul~\citep{jitvul} by 32.7 points in P-C; although JitVul navigates the same interprocedural call graph through caller/callee tools, its verdicts are not grounded in verified evidence, which isolates the contribution of our CPG-verified reward from that of tool access alone. \sys further exceeds its SFT-only baseline by 11.4 points in P-C while reducing the mean number of tool calls per rollout from 7.6 to 4.3, so the gain comes from shorter, better-targeted investigations rather than from more search; since the policy is a 7B open model, this accuracy is obtained at a fraction of the inference cost of the frontier baselines it surpasses. Its advantage persists on an out-of-distribution corpus and under class imbalance: because real code is overwhelmingly benign, we also evaluate at vulnerable:benign ratios from 1:2 to 1:16, and \sys leads at every ratio, attaining 0.275 F1 against 0.197 for the SFT baseline at 1:8.

In summary, our main contributions are as follows: 1) We propose \sys, an agentic RL framework in which a single CPG serves as both the policy's queryable tool and the reward's verifier for interprocedural vulnerability detection. 2) We devise hint-guided teacher distillation combined with a CPG-verified composite reward for stable agentic RL, and show that the SFT warm-start is structurally required, a finding of independent interest to practitioners. 3) We empirically show that 71.7\% of real vulnerable functions require interprocedural evidence, and demonstrate state-of-the-art pair-wise correctness over frontier and agent baselines under a leakage-free repository-level split.

\section{Problem Statement and Motivation}\label{sec:motivation}

In this section, we formulate the task of evidence-grounded agentic vulnerability detection and empirically analyze two limitations of existing approaches: 1) interprocedural context is unreachable at the function level; and 2) outcome-only rewards degenerate to evidence-free prediction.

\subsection{Problem Definition}\label{sec:problem}

Given a function $X$ and the Code Property Graph $G$ of the project that contains it, the task is to produce a structured report $Y$:
\begin{equation}
Y = \langle\, \ell,\ c,\ \mathcal{E}\,\rangle, \qquad
\ell \in \{\text{safe},\,\text{vulnerable}\}
\label{eq:report}
\end{equation}
where $\ell$ is the verdict, $c$ is a CWE class identifier if $\ell = \text{vulnerable}$ (else $c = \emptyset$), and $\mathcal{E} \subseteq V(G)$ is a set of CPG nodes that constitute the cited evidence. Crucially, the model is permitted to issue $T$ tool calls $\{q_1, \ldots, q_T\}$ on $G$ during inference, each returning a sub-graph $G_{q_t} \subseteq G$; the report $Y$ is produced after this interaction rather than from $X$ alone. At training time, ground truth $(\ell^*, c^*, \mathcal{E}^*)$ is derived from a frontier teacher LLM executed on the same $(X, G)$ with the same tool interface.

\subsection{Interprocedural Context Is Unreachable at the Function Level}
\label{sec:context_gap}

To measure whether the function body alone suffices for a correct verdict, we sample 138 vulnerable functions from our corpus and employ a frontier-LLM judge (Claude Opus 4.8). Given the function body and a summary of the confirmed defect, the judge classifies each function by whether the verdict is decidable from the body alone, or whether it requires evidence from at least one caller, callee, or dataflow predecessor \emph{outside} the function. We observe that 99 of 138 functions (71.7\%) require interprocedural evidence; among these cases, the required evidence most often resides within the implementation of a callee (60\%), followed by caller-side validation (19\%), dataflow predecessors (8\%), and external type definitions or global constants (13\%). \Cref{fig:context-example} illustrates this limitation. The function \emph{pj\_scan\_get\_char} from PJSIP is the site of CVE-2022-21723, a heap out-of-bounds read (CWE-125) caused by a missing bound check. In isolation the body reads as a benign character access: exploitability depends on buffer bounds defined in an external structure and on the \emph{caller}, whose malformed-multipart parsing advances the read pointer past the buffer end. The evidence therefore resides one type definition and one caller away. We therefore equip the model with a queryable CPG tool interface so that it can request exactly the context a verdict requires, on demand.

\begin{figure}[t]
\begin{lstlisting}[emph={pj_scan_get_char,pj_scan_syntax_err,pj_scan_skip_whitespace,PJ_SCAN_IS_PROBABLY_SPACE,PJ_DEF,pj_scanner}, moredelim={[is][\bfseries]{|}{|}}]
PJ_DEF(int) pj_scan_get_char(pj_scanner *scanner)
{
    |int chr = *scanner->curptr;|   // [!] unchecked read past buffer end
    if (!chr) {
        pj_scan_syntax_err(scanner);
        return 0;
    }
    ++scanner->curptr;
    if (PJ_SCAN_IS_PROBABLY_SPACE(*scanner->curptr)
        && scanner->skip_ws)
        pj_scan_skip_whitespace(scanner);
    return chr;
}
\end{lstlisting}
\caption{CVE-2022-21723 (PJSIP, CWE-125): the out-of-bounds read in \emph{pj\_scan\_get\_char} is undecidable from the function body alone.}
\label{fig:context-example}
\vspace{-5mm}
\end{figure}

\subsection{Evidence-Free Degeneration of Outcome-Only Rewards}
\label{sec:reward_degeneration}

To probe whether an outcome-only reward suffices for agentic training, we train GRPO policies under a reward that scores verdict correctness alone. We observe that the resulting policy drifts toward over-predicting ``vulnerable'': it labels both versions of a (vulnerable, patched) pair as vulnerable in 31.7\% of pairs, compared with 19.0\% under our full composite reward, and its pair-wise-correct rate reaches only 27.7\%, compared with 37.8\% under the composite reward (\Cref{sec:results}). The mechanism is structural: nothing in a verdict-only objective distinguishes a grounded verdict from a guess, so the policy can maximize reward by exploiting the class prior instead of investigating.

\begin{figure}[t]
\centering
\includegraphics[trim=0 60 0 45, clip, width=\columnwidth]{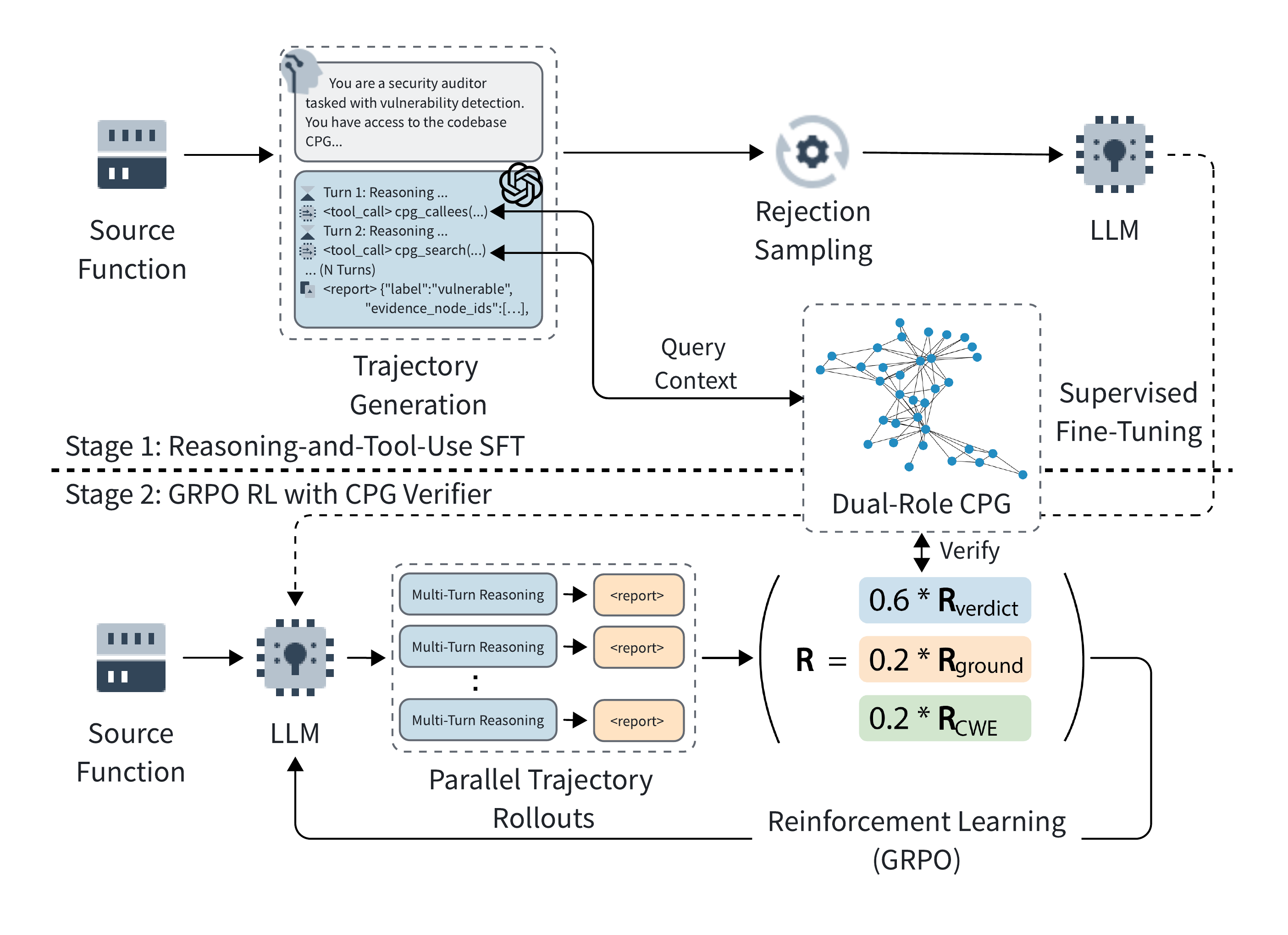}
\caption{Overview of the \sys training pipeline. (1) Stage 1: Reasoning-and-Tool-Use SFT. A teacher LLM investigates each function by querying the CPG and produces multi-turn trajectories, which are filtered via rejection sampling to form a high-quality corpus; the student model is then fine-tuned to learn the query-then-report format. (2) Stage 2: GRPO RL with CPG Verifier. The SFT model is refined with reinforcement learning: for each function the policy generates parallel multi-turn rollouts, which are scored by a composite reward over verdict correctness, evidence grounding, and CWE class, with the cited evidence nodes verified against the same CPG the policy queried.}
\label{fig:pipeline}
\vspace{-5mm}
\end{figure}

\section{Approach}\label{sec:approach}

Driven by the limitations analyzed in \Cref{sec:motivation}, we propose \sys, an agentic reinforcement-learning framework that learns to \emph{gather and cite} interprocedural evidence rather than predict a verdict from a single function alone. The design of \sys is guided by three core principles:

\noindent \textbf{1) Agentic Context Acquisition:} function-level detection cannot reach the interprocedural evidence on which most real-world verdicts depend (\Cref{sec:context_gap}). To close this gap, the policy issues typed queries over the project's CPG in a multi-turn loop, fetching exactly the callers, callees, dataflow paths, and other queries a verdict requires instead of classifying the function in isolation.

\noindent \textbf{2) Verifiable Evidence-Grounded Reward:} outcome-only rewards degenerate to evidence-free prediction (\Cref{sec:reward_degeneration}). We therefore score not only the final verdict but also the cited evidence and CWE class. Because Joern assigns every CPG node a persistent integer ID, the same graph the policy queries also verifies its citations by exact set comparison against the teacher, rendering the evidence signal deterministic and checkable.

\noindent \textbf{3) Distillation-Based Initialization:} agentic RL cannot bootstrap tool use from a cold start, since a policy that never issues tool calls gives the group-relative advantage no investigation rollouts to compare (\Cref{sec:results}). We therefore warm-start the policy by distilling hint-guided teacher trajectories before reinforcement learning.

These principles are realized through a two-stage training pipeline, summarized in \Cref{fig:pipeline}. In the first stage, we bootstrap the agent's tool-use and reporting behavior: a frontier teacher LLM, given the same CPG tool layer the student will use, produces multi-turn trajectories that we filter by rejection sampling and use for SFT. This stage yields a policy that reliably follows the query-then-report protocol. In the second stage, we refine that policy with GRPO under a composite reward whose evidence and CWE terms are verified against the teacher through the CPG. The final \sys model takes a function, interactively queries the project's CPG, and emits a structured report with a verdict, CWE class, and cited evidence nodes.

\subsection{Stage 1: Reasoning-and-Tool-Use SFT}

The first stage of our pipeline aims to initialize the model to follow a \emph{query-then-report} paradigm through knowledge distillation.

\subsubsection{Teacher Trajectory Distillation}
A frontier LLM (Claude Opus 4.7) is given access to the same CPG tool layer the student will use at inference. To fully exploit the teacher's capabilities, we define a structured investigation protocol that mirrors the workflow of a human security auditor: (1) \emph{Inspect}: identify parameters, sinks, and dangerous APIs in the function body; (2) \emph{Hypothesize}: propose threat models based on the observed patterns; (3) \emph{Query}: gather interprocedural evidence (sources, sinks, sanitizers, authorization checks) through the CPG tools; (4) \emph{Falsify}: for every vulnerability claim, issue queries that could disprove it, such as searching for sanitizers or validation on the suspected path; and (5) \emph{Decide}: emit the final report. A ``vulnerable'' verdict requires at least one falsification query before reporting. For each labeled (vulnerable, safe) function pair in our corpus, the teacher produces a multi-turn trajectory under this protocol: reasoning text interleaved with tool-call blocks that query the CPG, ending in a structured JSON report containing the verdict, predicted CWE class, cited evidence node IDs, and a free-text justification. Trajectories adhering to this protocol transfer the teacher's investigation behavior (which tools to call, in what order, and which nodes to cite) to the student.

\subsubsection{Hint-Guided Exploration}
A central design choice is how much of the ground truth the teacher is shown. Revealing the full answer (the CVE description or the patched lines) would yield post-hoc rationalizations: trajectories that justify a known conclusion rather than conduct an investigation. Withholding all information, conversely, expends most of the teacher budget on unsuccessful investigations. We adopt an intermediate design: on vulnerable functions the teacher receives only the CWE class identifier (e.g., CWE-22) as a directional hint, with no description, location, or evidence; safe functions carry no hint. The teacher must therefore locate the defect and assemble its evidence through its own CPG queries, so the cited evidence node IDs are the product of genuine tool use rather than a paraphrase of a revealed answer, which is precisely the property that qualifies them as grounding targets in \Cref{sec:reward}. The hint never reaches the student: the teacher is instructed not to reference it in its output, and the student's training prompts are rebuilt hint-free, byte-identical to the evaluation prompts, so no label information leaks into training.

\subsubsection{Two-Layer Rejection Sampling}
Teacher outputs are imperfect: rationales may be flawed and verdicts incorrect. Therefore, we adopt a two-layer rejection-sampling strategy to denoise the trajectories: 1) \emph{format and verdict filtering} discards any trajectory that violates the output schema (missing or malformed tool-call or report tags, unparseable JSON) or whose final verdict does not match the ground-truth label; 2) \emph{evidence-node validity filtering} retains a trajectory only if every cited node ID exists in the CPG and lies within the patched function's two-hop neighborhood, rejecting trajectories that reach the correct verdict while citing irrelevant or hallucinated evidence. Trajectories that pass both layers form the SFT corpus.

\subsubsection{SFT}
We fine-tune Qwen2.5-Coder-7B-Instruct~\citep{qwen} on the multi-turn trajectory corpus. Let $\tau_i$ denote a kept trajectory and $A_i$ the set of \emph{assistant} token positions in it (reasoning, tool calls, and the final report, excluding tool-return inputs). We minimize the masked negative log-likelihood over assistant tokens only:
\begin{equation}
\mathcal{L}_{\text{SFT}} = -\frac{1}{\sum_i |A_i|} \sum_i \sum_{t \in A_i} \log \pi_\theta\!\left(\tau_{i,t} \mid \tau_{i,<t}\right).
\label{eq:sft}
\end{equation}
The resulting policy, $\theta_{\text{SFT}}$, reliably issues CPG queries and emits well-formed reports, and serves as the initial policy for Stage~2.

\subsection{Stage 2: GRPO with CPG Verifier}
\label{sec:RL_Stage}

Following the SFT stage, \sys employs reinforcement learning with a CPG-verified reward, enabling the policy to surpass the teacher it imitates: maximizing likelihood against teacher trajectories provides no pressure to investigate more efficiently or to ground verdicts the teacher missed, whereas our composite reward credits grounded verdicts over unverified prediction.
We employ the on-policy algorithm GRPO~\citep{grpo} for this stage, with prompts sampled under a curriculum that shifts from shorter to longer functions as training proceeds. During training, for a sampled prompt $X$ with ground truth $(\ell^*, c^*, \mathcal{E}^*)$, the current policy $\pi_\theta$ rolls out $G$ complete multi-turn trajectories $\{\tau_1, \ldots, \tau_G\}$, each interleaving CPG tool calls with reasoning and ending in a final report. We parse each $\tau_i$ into a structured report $Y_i = \langle \ell_i, c_i, \mathcal{E}_i \rangle$ and compute a scalar reward against the teacher-derived ground truth. This reward is then used to update the policy, encouraging reports that agree with the ground truth in verdict, cited evidence, and CWE class.

\subsubsection{Reward Design}\label{sec:reward}
Our reward assesses the agreement between the generated report and the teacher-derived ground truth at three granularities (verdict, evidence, and CWE class), each a pure operation over labels or CPG node sets. Every component is \emph{format-gated}: a rollout with no parseable report scores 0. The reward is therefore entirely rule-based: there is no learned reward model, so the signal is deterministic, inexpensive to compute, and immune to the reward-model drift associated with learned critics.

\subsubsection{Verdict Agreement}
A binary indicator that the predicted verdict matches ground truth:
\begin{equation}
R_{\text{verdict}} = \mathbb{1}[\ell_i = \ell^*].
\end{equation}

\subsubsection{Evidence Grounding}
The $F_1$ overlap between the rollout's cited node set $S_{\text{pred}} = \mathcal{E}_i$ and the teacher's set $S_{\text{gt}} = \mathcal{E}^*$, over persistent CPG node IDs. With precision $P = |S_{\text{pred}} \cap S_{\text{gt}}| / |S_{\text{pred}}|$ and recall $R = |S_{\text{pred}} \cap S_{\text{gt}}| / |S_{\text{gt}}|$,
\begin{equation}
R_{\text{ground}} = \frac{2\,P\,R}{P + R}.
\end{equation}
We use $F_1$ rather than recall alone so the policy cannot inflate the score by citing many spurious IDs: precision penalizes over-citation. Because the IDs are addresses into the same graph the teacher cited from, this comparison is exact, with no textual fuzzy matching.

\subsubsection{CWE Agreement}
A tiered score, awarded only when both ground truth and prediction are ``vulnerable.'' It is $1.0$ for an exact CWE match, $0.5$ when the predicted class is a direct parent or child of the ground-truth class in the CWE-1000 hierarchy, and $0.0$ otherwise:
\begin{equation}
R_{\text{cwe}} =
\begin{cases}
1.0 & \text{if } c_i = c^*,\\
0.5 & \text{if } c_i \text{ is one hop from } c^* \text{ in CWE-1000},\\
0.0 & \text{otherwise,}
\end{cases}
\end{equation}
and is evaluated only for $\ell_i = \ell^* = \text{vuln}$ (else $R_{\text{cwe}} = 0$). The partial credit rewards predicting a closely related weakness (e.g., CWE-787 when the truth is its parent CWE-119) without crediting distant relatives.

\subsubsection{Composite Reward}
The final reward is a verdict-dominant weighted sum of the three terms:
\begin{equation}
R = 0.6\,R_{\text{verdict}} + 0.2\,R_{\text{ground}} + 0.2\,R_{\text{cwe}}.
\label{eq:reward}
\end{equation}
The $0.6$ verdict weight is deliberately dominant: a correct verdict alone (0.6) outscores a wrong verdict with maximal grounding and CWE (0.4), so the strongest gradient pressure is always on verdict correctness, while grounding and CWE act as smaller shaping signals that reward substantiated investigation. Because the weights sum to one and each component lies in $[0,1]$, the composite reward is itself bounded in $[0,1]$.

\subsubsection{GRPO Optimization}
Each of the $G$ rollouts for a prompt is scored to $R_i = R(Y_i, \ell^*, c^*, \mathcal{E}^*)$. GRPO uses the group as its own baseline: the advantage is the reward centered on the group mean,
\begin{equation}
A_i = R_i - \frac{1}{G}\sum_{j=1}^{G} R_j,
\label{eq:advantage}
\end{equation}
so a trajectory is favored when it scores above its siblings for the same function. The actor is then updated with the clipped surrogate objective using $A_i$ as the per-token advantage:
\begin{equation}
\mathcal{L}_{\text{GRPO}} = -\mathbb{E}_i\!\left[ \min\!\left( r_i A_i,\ \mathrm{clip}(r_i, 1{-}\epsilon, 1{+}\epsilon) A_i \right)\right],
\label{eq:grpo}
\end{equation}
where $r_i = \pi_\theta(a|s) / \pi_{\theta_{\text{old}}}(a|s)$ is the importance ratio and $\epsilon$ the clip threshold. The key property of this loop is that the reward computation queries the CPG $G$: evidence grounding is verified by the same graph the policy queried, closing the loop between the agent's tool and its verifier. The resulting tuned policy constitutes the final \sys model.

\section{Experimental Setup}\label{sec:setup}

In this section, we describe the baselines and evaluation metrics, the construction of our dataset, and the implementation details of \sys.

\subsection{Baselines and Evaluation Metrics}
We benchmark \sys against LLM-agent approaches (PrimeVul CoT~\citep{primevul}, VulTrial~\citep{vultrial}, and the state-of-the-art JitVul~\citep{jitvul}, whose ReAct agent navigates interprocedural context through caller/callee tools much like ours), frontier models (Claude Opus 4.7, GPT-5.5), and our own 0-shot and SFT-only models to isolate the contribution of the RL stage. For a controlled comparison, we run JitVul's ReAct agent on our own test set using its released prompts and tools, supplying the same interprocedural call graphs derived from our CPG.
We evaluate all models with overall accuracy, vulnerable-class precision, recall, and F1; for our own model variants we additionally report the mean number of tool calls per rollout. Following PrimeVul~\citep{primevul}, we adopt P-C as our primary metric: it credits a (vulnerable, patched) pair only when both of its functions are classified correctly, and is therefore robust to one-sided over-prediction.

\subsection{Dataset}\label{subsec:dataset}
To strictly prevent data leakage and evaluate cross-project generalization, we enforce a repository-level split where no repository overlaps between training, validation, and test sets. We adopt PrimeVul~\citep{primevul} as our primary corpus and construct an 8:1:1 train/validation/test split at the repository level: all functions from a given repository appear in exactly one of the three splits. Each example is a paired (safe, vulnerable) snapshot of the same function (one at the vulnerable commit, one at the patched commit), which forces the model to discriminate on the actual bug rather than on surface syntactic differences. We further apply a per-commit deduplication: when a fix commit modifies multiple functions, we retain only the one, neutralizing within-commit leakage where sibling functions share root cause, CWE, and CPG-reachable context. After this deduplication and the removal of functions for which Joern failed to produce a parseable CPG, our PrimeVul corpus contains 1,122 unique functions over 275 repositories.
To additionally evaluate how well our model generalizes to entirely unseen projects beyond PrimeVul, we construct a new out-of-distribution test set TitanVul\textsubscript{OOD}, derived from the vulnerability corpus TitanVul~\citep{titanvul}. To ensure a truly external test set, we filter out any function whose repository is present in any of our PrimeVul splits, and additionally restrict TitanVul\textsubscript{OOD} to C/C++ functions to match PrimeVul's language coverage. This process results in a clean out-of-distribution test set of 648 functions spanning 217 repositories.

\subsection{Implementation Details}
We implement \sys based on Qwen2.5-Coder-7B-Instruct~\citep{qwen} using Verl~\citep{verl}, with a Joern-based~\citep{joern} CPG service behind the tool layer. For SFT, we fine-tune for 5 epochs with a learning rate of $1\times10^{-4}$, computing loss exclusively on assistant spans (reasoning, tool calls, report). For RL, we employ GRPO initialized from the merged SFT checkpoint, with $G=8$ rollouts per prompt, a batch of 4 prompts (32 rollouts per step), a clip ratio of $\epsilon=0.2$, and an actor learning rate of $3\times10^{-6}$. All training runs on 4$\times$H100 80\,GB GPUs.

\section{Results}\label{sec:results}

We evaluate the effectiveness of \sys. The comparison with baselines is summarized in \Cref{tab:main}, out-of-distribution generalization in \Cref{tab:ood}, robustness under class imbalance in \Cref{tab:imbalance}, the progressive component ablation in \Cref{fig:ablation}, and the SFT-initialization study in \Cref{fig:training_curve}.

\begin{table}[t]
\caption{Comparison of \sys with state-of-the-art baselines and its own training stages on the held-out test set.}
\label{tab:main}
\centering
\small
\setlength{\tabcolsep}{3pt}
\begin{tabular}{l l c c c c c}
\toprule
\textbf{Approach} & \textbf{Strategy} & \textbf{P-C} & \textbf{Acc.} & \textbf{Prec.} & \textbf{Rec.} & \textbf{F1} \\
\midrule
Opus 4.7 & NA & 0.273 & 0.621 & 0.613 & 0.657 & 0.634 \\
GPT-5.5         & NA & 0.152 & 0.561 & 0.548 & 0.938 & 0.692 \\
PrimeVul & CoT         & 0.202 & 0.571 & 0.613 & 0.384 & 0.472 \\
VulTrial & Multi-Agent & 0.221 & 0.552 & 0.549 & 0.575 & 0.562 \\
JitVul   & ReAct Agent & 0.051 & 0.500 & 0.500 & 0.768 & 0.606 \\
\midrule
\sys          & NA     & 0.066 & 0.523 & 0.567 & 0.283 & 0.377 \\
\sys          & SFT    & 0.264 & 0.545 & 0.542 & 0.521 & 0.531 \\
\textbf{\sys} & SFT + RL & \textbf{0.378} & 0.633 & 0.659 & 0.567 & 0.603 \\
\bottomrule
\end{tabular}%
\end{table}

\subsubsection{1) \sys performs better than state-of-the-art baselines}
\Cref{tab:main} shows that \sys outperforms all baselines on P-C and accuracy.
P-C is a strict metric: a pair is counted as correct only when the model classifies \emph{both} the vulnerable function and its patched version correctly. A model that labels most functions as vulnerable can therefore obtain high per-function recall and F1 while obtaining low P-C. This is why P-C values are low for all methods.
\sys surpasses the best-performing baseline, the frontier model Claude Opus 4.7, by 10.5 points in P-C (0.378 vs.\ 0.273) and 1.2 points in accuracy (0.633 vs.\ 0.621) despite employing a 7B open model, and it likewise surpasses the multi-agent VulTrial (P-C 0.221) and the CoT PrimeVul baseline (P-C 0.202).
Notably, \sys also outperforms JitVul, the state-of-the-art agentic method for vulnerability detection, by 32.7 points in P-C (0.378 vs.\ 0.051). 
Moreover, \sys outperforms the SFT-only baseline with gains of 11.4 points in P-C (0.378 vs.\ 0.264) and 8.8 points in accuracy (0.633 vs.\ 0.545), while reducing the mean number of tool calls per rollout from 7.6 to 4.3, confirming the superiority of our training framework over standard SFT.
We further assess statistical significance with a paired bootstrap test: the improvement of \sys in P-C is significant against JitVul ($p < 0.001$), GPT-5.5 ($p < 0.001$), and PrimeVul CoT ($p = 0.001$), as well as against the strongest baseline, Claude Opus 4.7 ($p = 0.041$).

\begin{table}[t]
\caption{Generalization of \sys to the out-of-distribution test set TitanVul\textsubscript{OOD}.}
\label{tab:ood}
\centering
\small
\setlength{\tabcolsep}{3pt}
\begin{tabular}{l l c c c c c}
\toprule
\textbf{Approach} & \textbf{Strategy} & \textbf{P-C} & \textbf{Acc.} & \textbf{Prec.} & \textbf{Rec.} & \textbf{F1} \\
\midrule
\sys          & NA       & 0.093 & 0.502 & 0.502 & 0.472 & 0.487 \\
\sys          & SFT      & 0.248 & 0.554 & 0.573 & 0.423 & 0.487 \\
\textbf{\sys} & SFT + RL & \textbf{0.302} & 0.570 & 0.569 & 0.577 & 0.573 \\
\bottomrule
\end{tabular}%
\vspace{-5mm}
\end{table}

\subsubsection{2) \sys generalizes to an out-of-distribution corpus}
\Cref{tab:ood} evaluates \sys on TitanVul\textsubscript{OOD}, the out-of-distribution test set described in \Cref{subsec:dataset}, whose repositories are entirely disjoint from every PrimeVul split. \sys again outperforms both of its training stages: it attains a P-C of 0.302, compared with 0.248 for the SFT-only model and 0.093 for the 0-shot base, and the highest accuracy (0.570). Moreover, recall rises from 0.423 for the SFT-only model to 0.577 at comparable precision, 0.569 against 0.573, indicating that reinforcement learning improves the identification of vulnerabilities in previously unseen projects rather than merely recalibrating the decision threshold. This result demonstrates that the gains from CPG-grounded training are not specific to the training corpus but transfer to independently collected projects.

\begin{table}[t]
\caption{\sys robustness under class imbalance at vulnerable:benign ratios 1:$r$.}
\label{tab:imbalance}
\centering
\small
\setlength{\tabcolsep}{4.5pt}
\begin{tabular}{l l c c c c}
\toprule
\textbf{Ratio} & \textbf{Strategy} & \textbf{Acc.} & \textbf{Prec.} & \textbf{Rec.} & \textbf{F1} \\
\midrule
1:2  & NA & 0.528 & 0.313 & 0.354 & 0.332 \\
     & SFT    & 0.551 & 0.374 & 0.508 & 0.431 \\
     & SFT + RL & 0.640 & 0.468 & 0.548 & \textbf{0.504} \\
\addlinespace
1:4  & NA & 0.562 & 0.185 & 0.354 & 0.243 \\
     & SFT    & 0.559 & 0.230 & 0.508 & 0.316 \\
     & SFT + RL & 0.659 & 0.304 & 0.548 & \textbf{0.391} \\
\addlinespace
1:8  & NA & 0.583 & 0.096 & 0.333 & 0.149 \\
     & SFT    & 0.564 & 0.123 & 0.497 & 0.197 \\
     & SFT + RL & 0.675 & 0.181 & 0.575 & \textbf{0.275} \\
\addlinespace
1:16 & NA & 0.594 & 0.041 & 0.267 & 0.071 \\
     & SFT    & 0.569 & 0.068 & 0.517 & 0.120 \\
     & SFT + RL & 0.686 & 0.114 & 0.667 & \textbf{0.194} \\
\bottomrule
\end{tabular}%
\vspace{-5mm}
\end{table}

\subsubsection{3) \sys performs consistently better under class imbalance}
\Cref{tab:imbalance} shows that \sys outperforms both the 0-shot base and the SFT baseline at every imbalance ratio. 
Specifically, at 1:8, \sys attains 0.275 F1, compared with 0.197 for the SFT baseline and 0.149 for the 0-shot model. 
This result demonstrates that the advantage observed in Finding 1 is not an artifact of the balanced split but persists under realistic skew.

\subsubsection{4) Each technical design progressively contributes to the final performance}
\begin{figure}[t]
\centering
\small
\setlength{\tabcolsep}{3pt}
\begin{tabular}{l l c c c c}
\toprule
\textbf{Variant} & \textbf{Added Component} & \textbf{P-C} & \textbf{Acc.} & \textbf{F1} & \textbf{Tools} \\
\midrule
A0 & Base (0-Shot)        & 0.066 & 0.523 & 0.377 & 0.0  \\
A1 & {}+ SFT              & 0.264 & 0.545 & 0.531 & 7.62 \\
A2 & {}+ GRPO (Verdict)   & 0.277 & 0.567 & 0.577 & 5.62 \\
A3 & {}+ CWE              & 0.289 & 0.586 & 0.545 & 4.58 \\
A4 & {}+ Grounding (Full) & \textbf{0.378} & \textbf{0.633} & \textbf{0.603} & 4.26 \\
\bottomrule
\end{tabular}

\vspace{2.5mm}
\begin{tikzpicture}
\begin{axis}[
  width=\columnwidth,
  height=4.2cm,
  symbolic x coords={A0,A1,A2,A3,A4},
  xtick=data,
  ymin=0.46, ymax=0.66,
  ylabel={Test Accuracy},
  xlabel={Ablation Variant (Cumulative)},
  bar width=15pt,
  enlarge x limits=0.15,
  grid=major,
  grid style={dashed, gray!30},
  tick label style={font=\scriptsize},
  label style={font=\scriptsize},
  every node near coord/.append style={font=\tiny, /pgf/number format/.cd, fixed, fixed zerofill, precision=3},
]
\addplot[ybar, bar shift=0pt, fill=orange!70, draw=orange!90!black,
         nodes near coords]
  coordinates {(A0,0.523) (A1,0.545) (A2,0.567) (A3,0.586) (A4,0.633)};

\addplot[sharp plot, mark=*, mark size=1.8pt, thick, black,
         mark options={fill=black}]
  coordinates {(A0,0.523) (A1,0.545) (A2,0.567) (A3,0.586) (A4,0.633)};
\end{axis}
\end{tikzpicture}
\caption{Step-wise ablation of \sys across training stages (A0$\rightarrow$A4). We confirm that each design component contributes to the final performance.}
\label{fig:ablation}
\vspace{-5mm}
\end{figure}
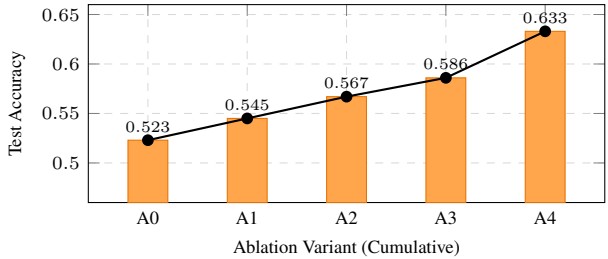

\sys trains the model in a multi-stage process. We therefore conduct a progressive ablation study in which each variant adds one component to the previous, as shown in \Cref{fig:ablation}. Specifically, we compare the following model variants: 1) Base Model (A0), the untrained base model with the tool interface; 2) SFT (A1), which adds Stage-1 supervised fine-tuning; 3) Verdict GRPO (A2), which adds GRPO with a verdict-only reward; 4) CWE-aware GRPO (A3), which adds the CWE-class term; and 5) the full \sys model (A4), which adds the evidence-grounding term to form the complete composite reward (\Cref{eq:reward}). Each variant is evaluated under the same protocol as above.
Every component improves performance. SFT raises accuracy from 0.523 to 0.545 and P-C from 0.066 to 0.264 by teaching the model to issue CPG queries and emit well-formed reports (A0$\rightarrow$A1). Optimizing the verdict-only reward raises accuracy to 0.567 and P-C to 0.277, learning which queries are decisive for the verdict (A1$\rightarrow$A2). Adding CWE-class credit raises accuracy by 1.9 points to 0.586 and P-C to 0.289 (A2$\rightarrow$A3). Finally, rewarding with the teacher's cited CPG nodes contributes the largest gain among the reward components, +4.7 points accuracy to 0.633 and +8.9 points P-C to 0.378, encouraging the policy to substantiate its verdicts rather than guess (A3$\rightarrow$A4). We confirm that each design progressively contributes to the final performance.

\subsubsection{5) SFT initialization is necessary}
\begin{figure}[t]
\centering
\begin{tikzpicture}
\begin{axis}[
  width=\columnwidth,
  height=5.0cm,
  xlabel={RL Training Step},
  ylabel={Per-Step Reward},
  xmin=1, xmax=30,
  ymin=0.10, ymax=0.55,
  grid=major,
  grid style={dashed, gray!30},
  tick label style={font=\scriptsize},
  label style={font=\scriptsize},
  legend style={font=\scriptsize, at={(0.02,0.98)}, anchor=north west,
                fill=white, fill opacity=0.9, draw=gray!50, nodes={inner sep=2pt}},
  legend cell align=left,
]
\addplot[draw=none, name path=up_sft, forget plot] coordinates {
  (1,0.2554) (2,0.2706) (3,0.2633) (4,0.2907) (5,0.2921) (6,0.3089) (7,0.3001)
  (8,0.3278) (9,0.3348) (10,0.3486) (11,0.3361) (12,0.3470) (13,0.3579)
  (14,0.3682) (15,0.3660) (16,0.3594) (17,0.3879) (18,0.3683) (19,0.3750)
  (20,0.4009) (21,0.4328) (22,0.4183) (23,0.4389) (24,0.4284) (25,0.4471)
  (26,0.4544) (27,0.4687) (28,0.4704) (29,0.4893) (30,0.4558)
};
\addplot[draw=none, name path=lo_sft, forget plot] coordinates {
  (1,0.1694) (2,0.1923) (3,0.1892) (4,0.2206) (5,0.2203) (6,0.2437) (7,0.2314)
  (8,0.2460) (9,0.2329) (10,0.2460) (11,0.2354) (12,0.2540) (13,0.2801)
  (14,0.3176) (15,0.3229) (16,0.3094) (17,0.3267) (18,0.3108) (19,0.3038)
  (20,0.3275) (21,0.3524) (22,0.3414) (23,0.3474) (24,0.3422) (25,0.3513)
  (26,0.3645) (27,0.3783) (28,0.3821) (29,0.3956) (30,0.3727)
};
\addplot[blue!40, fill opacity=0.5, forget plot] fill between[of=up_sft and lo_sft];
\addplot[draw=none, name path=up_0, forget plot] coordinates {
  (1,0.3062) (2,0.3008) (3,0.2931) (4,0.2994) (5,0.2994) (6,0.2894) (7,0.3000)
  (8,0.3319) (9,0.3344) (10,0.3381) (11,0.3381) (12,0.3381) (13,0.3150)
  (14,0.3125) (15,0.3144) (16,0.3156) (17,0.3044) (18,0.3031) (19,0.3031)
  (20,0.2975) (21,0.3000) (22,0.3038) (23,0.3038) (24,0.3038) (25,0.3113)
  (26,0.3075) (27,0.3163) (28,0.3313) (29,0.3391) (30,0.3396)
};
\addplot[draw=none, name path=lo_0, forget plot] coordinates {
  (1,0.2583) (2,0.2641) (3,0.2525) (4,0.2550) (5,0.2550) (6,0.2594) (7,0.2606)
  (8,0.2756) (9,0.2756) (10,0.2875) (11,0.2725) (12,0.2675) (13,0.2638)
  (14,0.2631) (15,0.2625) (16,0.2775) (17,0.2700) (18,0.2700) (19,0.2594)
  (20,0.2556) (21,0.2556) (22,0.2625) (23,0.2675) (24,0.2856) (25,0.2894)
  (26,0.2856) (27,0.2938) (28,0.2963) (29,0.2953) (30,0.2938)
};
\addplot[orange!50, fill opacity=0.5, forget plot] fill between[of=up_0 and lo_0];
\addplot[blue!70!black, very thick] coordinates {
  (1,0.2124) (2,0.2314) (3,0.2263) (4,0.2557) (5,0.2562) (6,0.2763) (7,0.2658)
  (8,0.2869) (9,0.2839) (10,0.2973) (11,0.2857) (12,0.3005) (13,0.3190)
  (14,0.3429) (15,0.3445) (16,0.3344) (17,0.3573) (18,0.3395) (19,0.3394)
  (20,0.3642) (21,0.3926) (22,0.3799) (23,0.3931) (24,0.3853) (25,0.3992)
  (26,0.4094) (27,0.4235) (28,0.4263) (29,0.4425) (30,0.4143)
};
\addlegendentry{SFT-Initialized}
\addplot[orange!90!black, very thick, densely dashed] coordinates {
  (1,0.2823) (2,0.2824) (3,0.2728) (4,0.2772) (5,0.2772) (6,0.2744) (7,0.2803)
  (8,0.3038) (9,0.3050) (10,0.3128) (11,0.3053) (12,0.3028) (13,0.2894)
  (14,0.2878) (15,0.2884) (16,0.2966) (17,0.2872) (18,0.2866) (19,0.2813)
  (20,0.2766) (21,0.2778) (22,0.2831) (23,0.2856) (24,0.2947) (25,0.3003)
  (26,0.2966) (27,0.3050) (28,0.3138) (29,0.3172) (30,0.3167)
};
\addlegendentry{Qwen2.5-7B-Initialized}
\end{axis}
\end{tikzpicture}
\caption{Training curves of the GRPO stage: SFT-initialized vs.\ 0-shot cold start.}
\label{fig:training_curve}
\vspace{-5mm}
\end{figure}
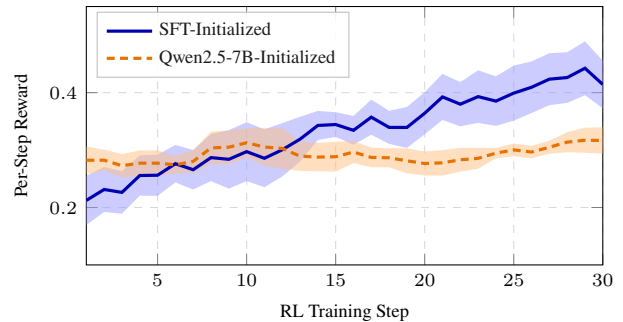

We investigate whether the SFT stage provides a necessary starting point for reinforcement learning. We compare two RL training runs: one initialized from our merged SFT checkpoint, and another that starts GRPO directly from the base Qwen2.5-Coder-7B-Instruct model with no SFT warm-up. \Cref{fig:training_curve} plots the learning curves for both settings. The cold-start policy remains at 1.0 turns per rollout throughout training, issuing zero tool calls in every rollout, compared with 16.6 turns for the SFT-initialized run, and it never acquires tool-use behavior. The mechanism is structural: with no investigation rollouts to compare within a group, the group-relative advantage provides no signal that would steer the policy toward querying. GRPO can refine an existing tool-use distribution but cannot induce one from a policy that never issues tool calls. Consequently, the SFT-initialized reward climbs steadily as training proceeds, whereas the cold-start reward exhibits almost no growth.

\section{Related Work}\label{sec:related}

\subsubsection{Static Analysis for Vulnerability Detection}
\citet{cpg} introduced the Code Property Graph, a unified program representation combining syntax, control flow, and data dependence. Static analyzers built on such representations, including Joern~\citep{joern} and CodeQL~\citep{codeql}, detect vulnerabilities through hand-authored dataflow queries: precise when a rule applies, but their rule sets are authored per language and are expensive to maintain. CPGs have also served as input to GNN-based detectors~\citep{devign}, but not, to our knowledge, as a \emph{tool surface} for an agentic LLM with the graph itself acting as the reward verifier, the dual role central to \sys (\Cref{sec:approach}). \sys reasons over the same substrate but replaces hand-authored rules with a learned query policy.

\subsubsection{LLM-Based Vulnerability Detection}
LineVul~\citep{linevul}, VulBERTa~\citep{vulberta}, Devign~\citep{devign}, and PrimeVul~\citep{primevul} treat detection as binary classification over a single function or hunk. They achieve strong intra-function accuracy but cannot reason across function boundaries, the regime in which most real CVEs arise (\Cref{sec:context_gap}). \sys extends this line by giving the model an explicit tool surface for interprocedural context rather than a fixed single-function input.

\subsubsection{Agentic RL and Process Supervision}
SWE-RL~\citep{swerl}, Agent~Q~\citep{agentq}, and DeepSeek-R1~\citep{grpo} train policies with verifiable end-task rewards; DeepSeek-R1 explicitly argues against process rewards, citing reward-hacking risk, while OpenAI's process reward model~\citep{prm} takes the opposite stance. \sys occupies an intermediate position: the verdict-match term is an end-task verifiable reward, while the evidence-grounding term is process supervision verified by the same CPG the model queried, with grounding ground truth derived from a teacher trajectory rather than an LLM judge.

\section{Conclusion and Future Work}\label{sec:conclusion}

We presented \sys, an agentic reinforcement-learning framework in which a Code Property Graph serves a dual role: the policy's queryable tool at inference time and the verifier of cited evidence at training time. Trained by distilling hint-guided teacher trajectories and refining the policy with a CPG-verified composite reward, a 7B model learns to gather interprocedural context and substantiate its verdicts with checkable evidence. It attains the best pair-wise-correct rate (0.378) and accuracy (0.633) among all compared approaches, surpassing frontier models restricted to the function body while issuing fewer tool calls than its SFT baseline, and its advantage persists out of distribution and under class imbalance. One finding is of independent interest to agentic-RL practitioners: the SFT warm-start is structurally required, since GRPO cannot bootstrap tool-use behavior it never samples. Because the CPG substrate is language-agnostic, cross-language evaluation and extension from detection to graph-verified repair are natural next steps. We release the pipeline, datasets, and per-step trajectories to support future agentic security-LLM research.

\newpage
\bibliography{refs}

\begin{thebibliography}{16}
\providecommand{\natexlab}[1]{#1}

\bibitem[{{Bytedance Research}(2024)}]{verl}
{Bytedance Research}. 2024.
\newblock {verl}: Volcano Engine Reinforcement Learning for {LLMs}.
\newblock \url{https://github.com/volcengine/verl}.

\bibitem[{Ding et~al.(2025)Ding, Fu, Ibrahim, Sitawarin, Chen, Alomair, Wagner, Ray, and Chen}]{primevul}
Ding, Y.; Fu, Y.; Ibrahim, O.; Sitawarin, C.; Chen, X.; Alomair, B.; Wagner, D.; Ray, B.; and Chen, Y. 2025.
\newblock Vulnerability Detection with Code Language Models: How Far Are We?
\newblock In \emph{Proceedings of the IEEE/ACM 47th International Conference on Software Engineering}.

\bibitem[{Fu and Tantithamthavorn(2022)}]{linevul}
Fu, M.; and Tantithamthavorn, C. 2022.
\newblock Linevul: A transformer-based line-level vulnerability prediction.
\newblock In \emph{Proceedings of the 19th international conference on mining software repositories}, 608--620.

\bibitem[{{GitHub}(2024)}]{codeql}
{GitHub}. 2024.
\newblock {CodeQL}.
\newblock \url{https://codeql.github.com}.

\bibitem[{Guo et~al.(2025)Guo, Yang, Zhang, Song, Wang, Zhu, Xu, Zhang, Ma, Bi et~al.}]{grpo}
Guo, D.; Yang, D.; Zhang, H.; Song, J.; Wang, P.; Zhu, Q.; Xu, R.; Zhang, R.; Ma, S.; Bi, X.; et~al. 2025.
\newblock DeepSeek-R1 incentivizes reasoning in LLMs through reinforcement learning.
\newblock \emph{Nature}, 645(8081): 633--638.

\bibitem[{Hanif and Maffeis(2022)}]{vulberta}
Hanif, H.; and Maffeis, S. 2022.
\newblock Vulberta: Simplified source code pre-training for vulnerability detection.
\newblock In \emph{2022 International joint conference on neural networks (IJCNN)}, 1--8. IEEE.

\bibitem[{Hui et~al.(2024)Hui, Yang, Cui, Yang, Liu, Zhang, Liu, Zhang, Yu, Lu et~al.}]{qwen}
Hui, B.; Yang, J.; Cui, Z.; Yang, J.; Liu, D.; Zhang, L.; Liu, T.; Zhang, J.; Yu, B.; Lu, K.; et~al. 2024.
\newblock Qwen2. 5-coder technical report.
\newblock \emph{arXiv preprint arXiv:2409.12186}.

\bibitem[{Li et~al.(2026)Li, Bui, Zhang, Weyssow, Yang, Zhou, Jiang, Chen, Huang, Nguyen et~al.}]{titanvul}
Li, Y.; Bui, N.~T.; Zhang, T.; Weyssow, M.; Yang, C.; Zhou, X.; Jiang, J.; Chen, J.; Huang, H.; Nguyen, H.~H.; et~al. 2026.
\newblock Out of Distribution, Out of Luck: How Well Can LLMs Trained on Vulnerability Datasets Detect Top 25 CWE Weaknesses?
\newblock In \emph{Proceedings of the IEEE/ACM 48th International Conference on Software Engineering}.

\bibitem[{Lightman et~al.(2024)Lightman, Kosaraju, Burda, Edwards, Baker, Lee, Leike, Schulman, Sutskever, and Cobbe}]{prm}
Lightman, H.; Kosaraju, V.; Burda, Y.; Edwards, H.; Baker, B.; Lee, T.; Leike, J.; Schulman, J.; Sutskever, I.; and Cobbe, K. 2024.
\newblock Let's verify step by step.
\newblock In \emph{International Conference on Learning Representations}, volume 2024, 39578--39601.

\bibitem[{Putta et~al.(2024)Putta, Mills, Garg, Motwani, Finn, Garg, and Rafailov}]{agentq}
Putta, P.; Mills, E.; Garg, N.; Motwani, S.; Finn, C.; Garg, D.; and Rafailov, R. 2024.
\newblock Agent Q: Advanced Reasoning and Learning for Autonomous AI Agents.
\newblock \emph{ArXiv}, abs/2408.07199.

\bibitem[{Wei et~al.(2026)Wei, Duchenne, Copet, Carbonneaux, Zhang, Fried, Synnaeve, Singh, and Wang}]{swerl}
Wei, Y.; Duchenne, O.; Copet, J.; Carbonneaux, Q.; Zhang, L.; Fried, D.; Synnaeve, G.; Singh, R.; and Wang, S. 2026.
\newblock Swe-rl: Advancing llm reasoning via reinforcement learning on open software evolution.
\newblock \emph{Advances in Neural Information Processing Systems}, 38: 78500--78525.

\bibitem[{Widyasari et~al.(2026)Widyasari, Weyssow, Irsan, Ang, Liauw, Ouh, Shar, Kang, and Lo}]{vultrial}
Widyasari, R.; Weyssow, M.; Irsan, I.~C.; Ang, H.~W.; Liauw, F.; Ouh, E.~L.; Shar, L.~K.; Kang, H.~J.; and Lo, D. 2026.
\newblock Let the Trial Begin: A Mock-Court Approach to Vulnerability Detection using LLM-Based Agents.
\newblock In \emph{Proceedings of the IEEE/ACM 48th International Conference on Software Engineering}.

\bibitem[{Yamaguchi et~al.(2014)Yamaguchi, Golde, Arp, and Rieck}]{joern}
Yamaguchi, F.; Golde, N.; Arp, D.; and Rieck, K. 2014.
\newblock Modeling and discovering vulnerabilities with code property graphs.
\newblock In \emph{2014 IEEE symposium on security and privacy}, 590--604. IEEE.

\bibitem[{Yamaguchi et~al.(2015)Yamaguchi, Maier, Gascon, and Rieck}]{cpg}
Yamaguchi, F.; Maier, A.; Gascon, H.; and Rieck, K. 2015.
\newblock Automatic inference of search patterns for taint-style vulnerabilities.
\newblock In \emph{2015 IEEE symposium on security and privacy}, 797--812. IEEE.

\bibitem[{Yildiz et~al.(2025)Yildiz, Teo, Lou, Feng, Wang, and Divakaran}]{jitvul}
Yildiz, A.; Teo, S.~G.; Lou, Y.; Feng, Y.; Wang, C.; and Divakaran, D.~M. 2025.
\newblock Benchmarking llms and llm-based agents in practical vulnerability detection for code repositories.
\newblock In \emph{Proceedings of the 63rd Annual Meeting of the Association for Computational Linguistics (Volume 1: Long Papers)}, 30848--30865.

\bibitem[{Zhou et~al.(2019)Zhou, Liu, Siow, Du, and Liu}]{devign}
Zhou, Y.; Liu, S.; Siow, J.; Du, X.; and Liu, Y. 2019.
\newblock Devign: Effective vulnerability identification by learning comprehensive program semantics via graph neural networks.
\newblock \emph{Advances in neural information processing systems}, 32.

\end{thebibliography}

\end{document}